\pdfoutput=1
\def\year{2019}\relax
\documentclass[letterpaper]{article} 
\usepackage{aaai19}  
\usepackage{times}  
\usepackage{helvet}  
\usepackage{courier}  
\usepackage{url}  
\usepackage{graphicx}  
\usepackage[ruled, linesnumbered]{algorithm2e}
\usepackage{amsmath}
\usepackage{xcolor}
\usepackage{multirow}
\usepackage{graphicx}
\usepackage{subfigure}
\usepackage{algorithmic}

\usepackage{amsfonts}

\begin{document}
%
\title{SocialGCN: An Efficient Graph Convolutional Network \\based Model for Social Recommendation}


\author{
	Le Wu\\
	Hefei University of Technology\\
	\texttt{lewu@hfut.edu.cn}
	\\\And
	Peijie Sun\\
	Hefei University of Technology\\
	\texttt{sun.hfut@gmail.com}
	\\\And
	Richang Hong\\
	Hefei University of Technology\\
	\texttt{hongrc.hfut@gmail.com}
	\\\AND
	Yanjie Fu\\
	Missouri University of Science and Technology\\
	\texttt{fuyan@mst.edu}
	\\\And
	Xiting Wang\\
	Microsoft Research Asia\\
	\texttt{xitwan@microsoft.com}
	\\\And
	Meng Wang\\
	Hefei University of Technology\\
	\texttt{eric.mengwang@gmail.com}
}

\maketitle

\begin{abstract}

Collaborative Filtering (CF) is one of the most successful approaches for recommender systems. With the emergence of online social networks, social recommendation has become a popular research direction. Most of these social recommendation models utilized each user's local neighbors' preferences to alleviate the data sparsity issue in CF. However, they only considered the local neighbors of each user and neglected the process that users' preferences are influenced as information diffuses in the social network. Recently, Graph Convolutional Networks~(GCN) have shown promising results by modeling the information diffusion process in graphs that leverage both graph structure and node feature information. To this end, in this paper, we propose an effective graph convolutional neural network based model for social recommendation. Based on a classical CF model, the key idea of our proposed model is that we borrow the strengths of GCNs to capture how users' preferences are influenced by the social diffusion process in social networks. The diffusion of users' preferences is built on a layer-wise diffusion manner, with the initial user embedding as a function of the current user's features and a free base user latent vector that is not contained in the user feature. Similarly, each item's latent vector is also a combination of the item's free latent vector, as well as its feature representation. Furthermore, we show that our proposed model is flexible when user and item features are not available.
	Finally, extensive experimental results on two real-world datasets clearly show the effectiveness of our proposed model.


\end{abstract}

\begin{small}
\begin{figure*} [htb]
	\begin{center}
		\includegraphics[width=170mm]{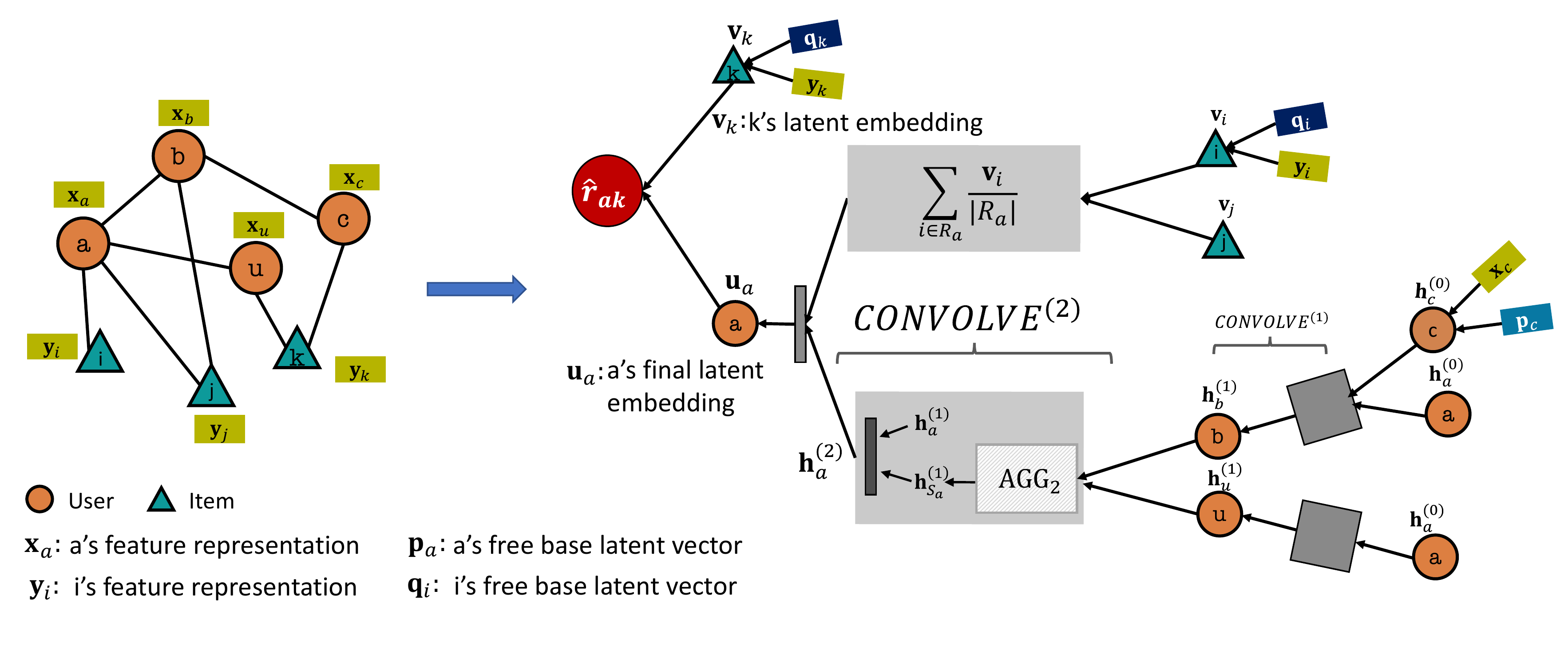}
	\end{center}
	\vspace{-1.0cm}
	\caption{The overall architecture of SocialGCN, where the left part shows a social recommendation platform and the right part depicts our proposed model.  As shown in this figure, to model the social diffusion process, for user $a$, her latent embedding in layer-2 is influenced from her social neighbors'~(i.e., $b$ and $u$) embeddings in layer-1 with graph convolutional operation. And each social neighbor's~(e.g., $b$) embedding in layer-1 is influenced from her respective neighbors~($c$ and $a$) embedding in layer-0.}\label{fig:gra_socialgcn}
	\vspace{-0.3cm}
\end{figure*}
\end{small}


\vspace{-0.2cm}
\section{Introduction}
\vspace{-0.1cm}

Collaborative Filtering~(CF) infers users' interests by modeling users' historical behaviors to items and is one of the most popular approaches for building recommender systems~\cite{su2009survey}. Among all CF models, latent-factor based approaches have received great success in both academia and industry due to their relatively high performance~\cite{koren2009matrix,UAI2009bpr,NIPS2008probabilistic}.  Specifically, given a user-item interaction matrix with sparse feedbacks, latent factor based models assumed each user and each item could be represented in a latent embedding space. Then, the predicted preference of a user to an item is reduced to comparing their embeddings in the latent space. 

With the prevalence of online social networks, more and more people like to express their opinions of items on these social platforms.  The social recommender systems have emerged as a promising direction, which leverage the social network among users to alleviate the data sparsity issue and improve recommendation performance~\cite{WSDM2011recommender,TKDE2014scalable,guo2015trustsvd,TKDE2014scalable}.  These approaches are based on the social influence assumption that connected people would influence each other, leading to the similar interests among social connections. E.g., social regularization has been empirically proven effective for social recommendation, with the assumption that connected users would share similar latent preferences ~\cite{Recsys2010matrix,WSDM2011recommender,TKDE2014scalable}. TrustSVD++ is proposed to incorporate the social neighbors' feedbacks to items as the auxiliary feedback of the active user~\cite{guo2015trustsvd}.  All these works empirically showed improvement with the social network modeling process. Nevertheless, nearly all these models leveraged the social network structure in a naive way by considering the local neighbors of each user. In a social network, as information would propagate from each user to her social neighbors, and then the social neighbors' neighbors, leading to an information diffusion process. Instead of the one-hop local social network structure, how to 
capture the influence diffusion process among users for better social recommendation performance? Furthermore, in most social platforms, users and items are associated with rich attributes, is it possible to for the designed model to be flexible to leverage the rich attributes of users and items?

By treating the social network as a graph structure, recent years  have shown significant improvement in learning embeddings of graph-structural data that could well tackle graph based tasks~\cite{IDEB2017representation}. A most prominent technique is Graph Convolutional Networks~(GCN), which shows theoretical elegance and relatively high performance in many graph-based tasks~\cite{IDEB2017representation,ICLR2017semi,ICLR2017graph}.  The key idea of GCNs is to learn the iterative convolutional operation in graphs, where each convolutional operation means generating the current node representations from the aggregation of local neighbors in the previous layer. Starting from a bottom layer of node representations as their node features, GCN stacks multiple convolutional operations to simulate the message passing of graphs. Therefore, both the information propagation process with graph structure and node attributes are well leveraged in GCNs. Recently, GCNs have also been explored in the recommender systems. Researchers proposed to transform the recommendation task as a link prediction problem in graphs and learned user and item latent embeddings through message passing on the bipartite user-item interaction graph~\cite{ICLR2017graph,KDD2018graph}. 
These models made preliminary attempts of adopting GCNs for recommendation with encouraging results. As the social network naturally models the influence propagation process of users' interests, we argue, is it possible to leverage the  social network and users' preferences under GCNs for social recommendation?


In this paper, we propose an effective graph convolutional neural network based model, i.e., SocialGCN,  for social recommendation.  The overall framework of SocialGCN is shown in Fig.\ref{fig:gra_socialgcn}.
Similar as many classical latent factor based models,  we assume the predicted preference is modeled as the inner product between user embeddings and items embeddings. Instead of shallow latent factor based models that directly learn the user embeddings and item embeddings, our key contribution lies in designing deep models that will capture the unique characteristics of social networks for user embedding and item embedding modeling. Specifically, we borrow the strengths of GCNs to capture how users' preferences are influenced by the social diffusion process in social networks. The diffusion of users' preferences is built on a layer-wise diffusion manner, with the initial user embedding as a function of the current user's features and a free base user latent vector that is not contained in the user feature. Similarly, each item's latent vector is also a combination of the item's free latent vector, as well as its feature representation. We further show that the proposed SocialGCN is flexible to apply to the scenario when user and item attributes are not available. In summary, to the best of our knowledge, we are one of the first few attempts to apply GCNs to model the social diffusion process for social recommendation. In the experimental results, SocialGCN outperforms more than \textbf{8}\% and \textbf{11}\% all metrics of \textit{Yelp} and \textit{Flickr} respectively.

\vspace{-0.1cm}
\section{Related Work}
\vspace{-0.1cm}


\textbf{Collaborative Filtering.}
Given an user-item rating  matrix {\small$\mathbf{R}$}, CF usually projected both users and items in a same low latent space. Then, each user's predicted preference of to an item could be measured by the similarity of the user's latent vector and the item latent vector in the learned low latent space~\cite{koren2009matrix,NIPS2008probabilistic}. In reality, compared to the explicit ratings, it is more common for users implicitly express their feedbacks through action or inaction, such as click, add to cart or consumption ~\cite{hu2008collaborative,UAI2009bpr}. Bayesian Personalized Ranking~(BPR) is a state-of-the-art latent factor based technique for dealing with implicit feedback. Instead of directly predicting each user's point-wise explicit ratings, BPR modeled the pair-wise preferences with the assumption that users prefer the observed implicit feedbacks compared to the unobserved ones~\cite{UAI2009bpr}. Despite the relatively high performance, cold-start problems are a barrier to the performance of these collaborative filtering models. To tackle the data sparsity issue, many models have been proposed by extending these classical CF models. E.g., SVD++ is proposed to combine users' implicit feedbacks and explicit feedbacks for modeling users' latent interests~\cite{KDD2008factorization}.  Besides, as users and items are associated with rich attributes, Factorization Machine~(FM) is such a unified model that leverages the user and item attributes in latent factor based models~\cite{ICDM2010factorization}.

\textbf{Social Recommendation}  With the prevalence of online social platforms, social recommendation has emerged as a promising direction that leverages the social network among users to enhance recommendation performance~\cite{WSDM2011recommender,guo2015trustsvd,TKDE2014scalable}. In fact, social scientists have long converged that as information diffuses in the social networks, users are influenced by their social connections with the social influence theory, leading to the phenomenon of similar preferences among social neighbors~\cite{KDD2008influence,ibarra1993power,Nature2012,KDD2018deepinf}. Social regularization has been empirically proven effective for social recommendation, with the assumption that similar users would share similar latent preferences under the popular latent factor based models~\cite{Recsys2010matrix,WSDM2011recommender}.
SBPR model is proposed into the pair-wise BPR model with the assumption that users tend to assign higher ratings to the items their friends prefer~\cite{CIKM2014leveraging}. By treating the social neighbors' preferences as the auxiliary implicit feedbacks of an active user, TrustSVD is proposed to incorporate the trust influence from social neighbors on top of SVD++~\cite{KDD2008factorization}. The proposed model has state-of-the-art better performance than previous social recommendation models~\cite{guo2015trustsvd,TKDE2016novel}. As items are associated with attribute information~(e.g., item description, item visual information), ContextMF is proposed to combine social context and social network under a collective matrix factorization framework with carefully designed regularization terms~\cite{TKDE2014scalable}.  In summary, all these social recommendation based models have shown the superior performance with the social network modeling. Nevertheless, current models were based on shallow models for leveraging the social network structure~(e.g., social regularization or combining social neighbors' preferences as auxiliary feedbacks). Instead of considering the
social neighbor information, our work differs from these works in explicitly modeling the users' latent preferences with information diffusion process in the social network.


\textbf{Graph Convolutional Networks.} Convolutional neural network has been proven successful in a diverse range of domains, such as images~\cite{krizhevsky2012imagenet} and text~\cite{kim2014convolutional}.
Compared to images and text that are lied in the regular domain, recently, research interests have been paid in generalizing convolutions to graphs, which are irregular in nature~\cite{IDEB2017representation}. Graph convolutional networks are of particular interests due to the theoretical elegance and their relatively high performance~\cite{IDEB2017representation,ICLR2017semi,ICLR2017graph}. The key idea of GCNs is to generate node embeddings in a message passing or information diffusion manner of a graph.  Specifically, each node obtains its embedding by aggregating information from the neighbors, and in turn, the message coming from the neighbors are based on the neighbors from their respective neighbors, and so on. These models are termed with convolution as the operation of aggregating from neighbors resembles the convolutional layer in computer vision~\cite{ICLR2017semi}.
GraphSAGE extended GCN to the inductive setting by learning a function that generates embeddings by sampling and aggregating features from a node's local neighbors~\cite{ICLR2017graph}. By extending the success of GCNs in graphs, researchers proposed to learn latent embeddings of users and items through message passing on the bipartite user-item interaction graph under~\cite{ICLR2017graph}. Researchers also developed a data-efficient GCN algorithm PinSage, which combines efficient random walks and graph convolutions to generate embeddings of nodes that incorporate both graph structure as well as node feature information~\cite{KDD2018graph}.  These preliminary attempts to apply GCNs to recommender systems have simply transformed user-item interaction matrix into a graph and focused on the efficiency issue for recommendation. Our proposed model differs from these works as we focus on leveraging GCNs
to model the social diffusion process for better social recommendation performance.




\vspace{-0.1cm}
\section{The Proposed Model}
\vspace{-0.1cm}
In a social based recommender system, there are two sets of entities: a user set {\small$U$~($|U|\!=\!M$)}, and an item set {\small$V$~($|V|\!=\!N$)}. Users interact with items in this system. As the implicit feedbacks~(e.g., browse, consumption) are more common in recommender systems, we also consider the implicit feedback scenario\cite{UAI2009bpr}. Let the rating matrix {\small $\mathbf{R}\in\mathbb{R}^{M\times N}$} denote users' implicit feedback to items, with $r_{ai}\!=\!1$ if user $a$ is interested in item $i$, otherwise it equals 0. The social link matrix  {\small $\mathbf{S}\in\mathbb{R}^{M\times M}$} denotes the social connections among users in the social network. If user $a$ follows user $b$, $s_{ba}=1$, otherwise it equals 0. Then, each user $a$'s  ego social network, i.e., the social neighbors that $a$ follows, is the i-th column~($S_a$) of {\small $\mathbf{S}$}. Besides, each user $a$ is associated with real-valued attributes, denoted as $\mathbf{x}_a$ in user attribute matrix {\small $\mathbf{X}\in\mathbb{R}^{d1\times M}$}. Also, each item $i$ has an attribute vector $\mathbf{y}_i$ in item attribute matrix {\small $\mathbf{Y}\in\mathbb{R}^{d2\times N}$}.  Then, the social recommendation task asks that, given a rating matrix {\small$\mathbf{R}$} and a social network {\small$\mathbf{S}$}, and associated feature matrix {\small $\mathbf{X}$} and {\small $\mathbf{Y}$} of users and items, our goal is to predict each user's preferences to unknown items.

\textbf{\vspace{-0.5cm}}
\subsection{Model Architecture}
\vspace{-0.1cm}
In this part, we build a SocialGCN model that depicts the influence of propagation on social networks for social recommendation. Similar as many latent factor based models, our goal is to encode both users and items in a low latent embedding space, such that the similarity in the latent space approximates the preference of users to items.

Let {\small $\mathbf{U}\in\mathbb{R}^{D\times M}$} and {\small $\mathbf{V}\in \mathbb{R}^{D\times N}$} denote the embeddings of users and items in the latent space. Then, the predicted preference of user $a$ to item $i$, denoted as $\hat{r}_{ai}$, is computed as:

\vspace{-0.4cm}
\begin{equation}\label{eq:pred_r}
\hat{r}_{ai} = \mathbf{u}_a * \mathbf{v}_i,
\end{equation}
\vspace{-0.4cm}

\noindent where $\mathbf{u}_a$ denote the a-th column lookup in the user embedding matrix {\small$\mathbf{U}$}, and  $\mathbf{v}_i$ is item $i$'s latent embedding in the item embedding matrix {\small$\mathbf{V}$}.

In fact, the traditional latent factor based approaches are shallow models, where each user~(item) latent embedding is directly an embedding lookup from the parameters of the embedding matrices~\cite{UAI2009bpr,KDD2008factorization,ICDM2010factorization}. However, there are several limitations of the shallow latent factor based models. First, as users' feedbacks are usually very sparse, simply relying on the shallow embedding could not model the complex aspects that may determine each user's~(item's) latent vector $\mathbf{u}_i$~($\mathbf{v}_i$). Besides, users and items are associated with rich attributes. To leverage the user and item attributes, an intuitive idea is to adopt the Factorization Machines~(FM) that learn user and item bias with attributes~\cite{ICDM2010factorization}.
This formulation in FM neglected the correlation between the attributes and the latent embeddings, which may restrict model capacity. Furthermore, in social recommendation systems, users connect with each other and the influence propagation among users could largely influence the formulation of user embedding matrix {\small$\mathbf{U}$}. Nevertheless, most latent factor based models for social recommendation neglect the influence propagation in social networks, leading to inferior performance. To tackle these challenges, we focus on how to model user latent embedding matrix {\small $\mathbf{U}$} and item latent embedding matrix {\small $\mathbf{V}$} for social recommendation, where the influence propagation and associated attributes can be properly modeled. In the following, we first introduce the item embedding modeling step, followed by the user embedding modeling that involves influence propagation. Without confusion we use $a$, $b$, $c$ to denote users and
$i$, $j$, $k$ to denote items.

%

\textbf{Item Embedding.} For each item $i$, we assume its latent embedding $\mathbf{v}_i$ is a function of two parts: the item feature embedding $\mathbf{y}_i$ and a free base latent vector $\mathbf{q}_i$ from a free base latent matrix {\small$\mathbf{Q}\in \mathbb{R}^{L\times N}$}. Specifically, the free base latent vector {\small $\mathbf{Q}$} models the item aspects that could not be captured by the item feature matrix {\small $\mathbf{Y}$}. Then, each item $i$'s latent embedding $\mathbf{v}_i$  can be formulated as：

\vspace{-0.2cm}
\begin{equation} \label{eq:item_embed}
\mathbf{v}_i = \sigma(\mathbf{F}\times[\mathbf{q}_i, \mathbf{y}_i]),
\vspace{-0.2cm}
\end{equation}

\noindent  where $[\mathbf{q}_i, \mathbf{y}_i]$ denotes the concatenation of item $i$'s feature vector and its corresponding free embedding, $\mathbf{F}$  is a transformation matrix. Then, we feed the input into a fully-connected neural network with a non-linear transformation function $\sigma$ to get $\mathbf{v}_i$. Without confusion, in this paper, we omit the bias term in a fully-connected neural network for notational convenience.

\textbf{User Embedding.} For each user $a$, the composition of her latent embedding $\mathbf{u}_i$ is more complicated, as users would like to express and propagate their preferences, leading to latent preference diffusion in social network $\mathbf{S}$. Therefore, each user's latent preference is influenced by her social neighbors, and each social neighbor also influenced by the social neighbor's neighbors. As information diffuses in social networks, we borrow the key ideas of GCNs to model the influence diffusion effect for user embedding modeling. Given a social network {\small$\mathbf{S}$}, GCN aims to model each node embeddings from its social neighbors with a hierarchical multi-layer structure.
For each user $a$, let $\mathbf{h}^{k}_a$ denotes her latent embedding in the $k$-th layer. Given the latent embeddings of her social neighbors' at this layer, the graph convolutional operation defines $a$'s latent embedding at the $k+1$-th layer, i.e., $\mathbf{h}^{k+1}_a$ as:

\begin{flalign}\label{eq:pro_gcn}
&\mathbf{h}^{k+1}_{Sa}=AGG_k(\mathbf{h}^k_b, b\in S_a), \nonumber\\
&\mathbf{h}^{k+1}_a=CONVOLVE^{(k)}([\mathbf{h}^{k+1}_{S_a}, \mathbf{h}^k_a]),
\end{flalign}

\noindent where the first function $AGG_k$  denotes that each user $a$ aggregates the influences from her social neighbors' latent embeddings at the $k$-th layer with ${\mathbf{h}^k_b, b\in S_a}$. Many aggregation functions could be applied, such as average aggregation or max aggregation. Then, we feed the concatenated vector of $\mathbf{h}^{k+1}_{Sa}$, as well as $a$'s latent embedding $\mathbf{h}^k_a$ in this layer to a fully connected layer of neural network with function $CONVOLVE^{(k)}$. In practice, we use a non-linear function $ReLU$ with transform matrix $\mathbf{W}^k$ to realize the $CONVOLVE^{(k)}$ as:

\vspace{-0.2cm}
\begin{equation} \label{eq:relu_to_convolve}
\mathbf{h}^{k+1}_a =ReLU(\mathbf{W}^k\times[\mathbf{h}^{k+1}_{S_a}, \mathbf{h}^k_a]).
\end{equation}

After that, we get $a$'s embedding in the $k+1$-th layer. Starting from the $\mathbf{h}^0_a$ of each user, the layer-wise graph convolutional operation clearly models the influence propagation of users' preferences in the social network.

In the original GCN model, with the feature vector $\mathbf{x}_a$ of each user, the layer-0 embedding of each user is defined as her input features:

\vspace{-0.4cm}
\begin{equation} \label{eq:base_gcn}
\mathbf{h}^0_a =\mathbf{x}_a.
\end{equation}
\vspace{-0.4cm}

In social recommender systems, each user's  layer-0 embedding vector captures the base latent embedding that propagates in the social network. We argue that the feature matrix $\mathbf{X}$ could not well capture users' latent interests for information propagation. Therefore, similar as item embeddings, we also associate each user with a free base latent vector $\mathbf{p}_a$ from the user free base latent matrix {\small$\mathbf{P}\in\mathbb{R}^{L\times M}$}.  This free base latent matrix captures each user's latent interests that could not be modeled by user feature matrix {\small$\mathbf{X}$}. Then, instead of assuming each user's layer-0 embedding in Eq.\eqref{eq:base_gcn}, we model $\mathbf{h}^0_a$ as a function of her features $\mathbf{x}_a$ and her
free base latent vector $\mathbf{p}_a$:

\begin{equation} \label{eq:base_user}
\vspace{-0.2cm}
\mathbf{h}^0_a = \sigma(\mathbf{W}^0\times[\mathbf{x}_a,\mathbf{p}_a]).
\vspace{-0.1cm}
\end{equation}

Combining the layer-0 user embedding~(Eq.\eqref{eq:base_user}, and the influence diffusion process~(Eq.\eqref{eq:pro_gcn}), with a predefined layer depth $K$,  we model each user's final latent embedding as:

\begin{equation} \label{eq:user_embed}
\vspace{-0.1cm}
\mathbf{u}_a =\mathbf{h}^K_a+\sum_{i\in R_a} \frac{\mathbf{v}_i}{|R_a|},
\vspace{-0.2cm}
\end{equation}

where $R_a\!=\![i|i: r_{ai}\!=\!1]$ is the itemset that $a$ likes. In this equation, each user's final latent representation $\mathbf{u}_a$ is a combination of two parts: the embeddings from the social diffusion process as: $\mathbf{h}^K_a$, and the preferences from her historical behaviors as: $\sum_{i\in R_a} \frac{\mathbf{v}_i}{|R_a|}$. In fact, leveraging the historical feedbacks of users for user embedding part resembles the SVD++ model~\cite{KDD2008factorization}, which has shown better performance over the classical latent factor based models. Our proposed user latent embedding part advances SVD++ by leveraging the user features in the social diffusion process with carefully designed diffusion interest vector $\mathbf{h}^K_a$.

\begin{small}
	{\renewcommand\baselinestretch{0.5}\selectfont
		\begin{algorithm}[htb]
			\renewcommand{\algorithmicrequire}{\textbf{Input:}}
			\renewcommand\algorithmicensure {\textbf{Output:}}
			\caption{ \small{The learning algorithm of SocialGCN}\ \ \ \ \ \  }\label{alg:sig}
			\begin{algorithmic}[1]
				\REQUIRE Rating matrix {\small$\mathbf{R}$}, social matrix {\small$\mathbf{S}$}, diffusion depth $K$; \\
				\ENSURE Parameter set $\Theta$ ;\\
				
				\STATE Initialize model parameter set $\Theta$  with small random  values;\\
				
				\WHILE {Not converged}
				\FOR {Each user-item pair {\small $<a,i>$ } in the training data}
				\STATE Compute the item embedding $\mathbf{v}_i$~(Eq.\eqref{eq:item_embed});
				\STATE Compute the input user embedding $\mathbf{h}^0_a$ at layer 0~(Eq.\eqref{eq:base_user});
				\FOR {$k=1$ to $K$}
				\STATE Compute the latent preference diffusion $\mathbf{h}^k_a$~(Eq.\eqref{eq:pro_gcn});
				\ENDFOR
				\STATE Compute the user embedding vector $\mathbf{u}_a$(Eq.\eqref{eq:user_embed});
				\STATE Compute the predicted rating {\small $\small \hat{r}_{ai}$}~(Eq.\eqref{eq:pred_r});
				\FOR {Each parameter $\theta$ in {$[\Theta_1,\Theta_2]$}}
				\STATE Update $\theta=\theta-\eta \frac{\partial L}{\partial \hat{r}_{ai}}\frac{\partial \hat{r}_{ai}}{\partial \theta}$ ~(Eq.\eqref{eq:loss_r});
				\ENDFOR
				\ENDFOR
				\ENDWHILE
				\STATE Return {\small $\Theta_1\!=\![\mathbf{P,Q}]$} and parameters in $\Theta_2$.
			\end{algorithmic}
		\end{algorithm}
		\par}
\end{small}

\vspace{-0.3cm}
\subsection{Model Training}
As we focus on implicit feedbacks of users, similar to the widely used ranking based loss function in BPR~\cite{UAI2009bpr}, we also design a pair-wise ranking  based loss function for optimization:

\begin{small}
	\vspace{-0.4cm}
	\begin{equation}\label{eq:loss_r}
	\min\limits_{\Theta} \mathcal{L}(\mathbf{R},\mathbf{\hat{R}})=\sum_{a=1}^M\sum\limits_{(i,j)\in D_a } s(\hat{r}_{ai}-\hat{r}_{aj}) +\lambda||\Theta_1||^2
	\end{equation}
	\vspace{-0.3cm}
\end{small}

\noindent where $s(x)$ is a sigmoid function. {\small $\Theta\!=\![\Theta_1,\Theta_2]$}, with {\small $\Theta_1\!=\![\mathbf{P},\mathbf{Q}]$}, and
{\small $\Theta_2\!=\![\mathbf{F}, {[\mathbf{W}^k]}_{k=1}^K]$}.  $\lambda$ is a regularization parameter that controls the complexity of user and item free embedding matrices. {\small$D_a=\{(i,j)|i\in R_a\!\wedge\!j\in V-R_a\}$} denotes the pairwise training data for $a$ with {\small$R_a$} represents the itemset that $a$ positively shows feedback.

All the parameters in the above loss function are differentiable. In practice, we implement the proposed model with TensorFlow\footnote{https://www.tensorflow.org} to train model parameters with mini-batch Adam. The detailed training algorithm is shown in Alg.~\ref{alg:sig}. In practice, we could only observe positive feedbacks of users with huge missing unobserved values, similar as many implicit feedback works, for each positive feedback, we randomly sample 5 missing unobserved feedbacks as pseudo negative feedbacks at each iteration in the training process~\cite{leAAAI2015}.  As each iteration the pseudo negative samples change, each missing value gives very weak negative signal.

\vspace{-0.3cm}
\subsection{Model Analysis}
\vspace{-0.1cm}
In this subsection, we give a detailed analysis of the proposed model.

\textbf{Space complexity.} As shown in Eq.\eqref{eq:loss_r}, the model parameters are composed of two parts: the user and item free embeddings {\small$\Theta_1\!=\![\mathbf{P},\mathbf{Q}]$}, and the parameter set {\small $\Theta_2\!=\![\mathbf{F}, {[\mathbf{W}^k]}_{k=1}^K]$}. Since most latent factor based models~(e.g., BPR~\cite{UAI2009bpr}) need to store the embeddings of each user and each item, the space complexity of $\Theta_1$ is the same as classical latent factor based models and grows linearly with users and items. For parameters in {\small$\Theta_2$}, as they are shared among all users and items, this additional storage cost is a constant. Therefore, the space complexity of SocialGCN is the same as classical latent factor based models.

\textbf{Time complexity.} Since our proposed loss function resembles BPR with a pair-wise loss, we compare the time complexity of SocialGCN with BPR. As shown in Alg.~\ref{alg:sig}, the main additional time cost lies in the influence diffusion process~(Line 6 to Line 8). The diffusion process costs {\small$O(MKL)$}, where $M$ is the number of users, and  $K$ denotes the diffusion depth and $L$ denotes the average social neighbors of each user. Similarly, the additional time complexity of updating parameters(Line 12) is {\small$O(MKL)$}. Therefore, the additional time complexity is {\small$O(MKL)$}. In fact, as shown in the empirical findings as well as our experimental results, most GCN based models reach the best performance when $K$=2 or $K$=3. Also, the average social neighbors per user are limited with {\small $L\ll M$}. Therefore, the additional time complexity is acceptable and the proposed SocialGCN could be applied to real-world social recommender systems.

\textbf{Model generalization.} We construct the proposed model when user and item attributes are available. In fact, our model is also applicable to the scenario when there are no associated attributes of users and items. Under this circumstance, as shown in Eq.\eqref{eq:item_embed}, each item's latent embedding $\mathbf{v}_i$ degenerates to $\mathbf{q}_i$. Similarly, each user's latent embedding $\mathbf{u}_a$ is changed as the layer-0 embedding $\mathbf{h}^0=\mathbf{p}_a$~(Eq.\eqref{eq:base_user}). The whole learning process is the same as shown in Alg.~\ref{alg:sig}.

\vspace{-0.2cm}
\section{Experiments}

In this section, we conduct experiments to evaluate the performance of SocialGCN on two datasets. Specifically, we aim to answer the following two research questions: First, does SocialGCN outperforms the state-of-the-art baselines for the social recommendation task?  Second, what's the effectiveness of each part in the SocialGCN model, e.g., diffusion modeling, attributes modeling, and so on.



\begin{table}\vspace{-0.2cm}
	\centering
	\setlength{\belowcaptionskip}{5pt} %
	\caption{The statistics of the two datasets.}\label{tab:stat}
	\begin{tabular}{c|c|c}
		\hline
		Dataset&\textit{Yelp}&\textit{Flickr}\\
		\hline
		Users&17237&8358\\
		Items&38342&82120\\
		\hline
		Total Links&143765&187273\\
		Training Ratings&185869&282444\\
		Test Rating&18579&32365\\
		\hline
		Link Density&0.048\%&0.268\%\\
		Rating Density&0.028\%&0.004\%\\
		\hline
	\end{tabular}
\vspace{-0.6cm}
\end{table}

\vspace{-0.3cm}
\subsection{Experimental Settings}
\vspace{-0.1cm}
\begin{table*}
	\begin{small}
		\centering
		\caption{HR@10 and NDCG@10 comparisons for different dimension size $D$.}\label{tab:hr_ndcg_d}
		\begin{tabular}{|c|c|c|c|c|c|c|c|c|c|c|c|c|}
			\hline
			\multirow{3}*{Models} &\multicolumn{6}{|c|}{\textit{Yelp}}&\multicolumn{6}{|c|}{\textit{Flickr}} \\
			\cline{2-13}
			&\multicolumn{3}{|c|}{HR}&\multicolumn{3}{|c|}{NDCG}&\multicolumn{3}{|c|}{HR}&\multicolumn{3}{|c|}{NDCG}\\
			\cline{2-13}
			&$D$=16&$D$=32&$D$=64&$D$=16&$D$=32&$D$=64&$D$=16&$D$=32&$D$=64&$D$=16&$D$=32&$D$=64\\
			\hline
			\hline
			BPR&0.2443&0.2632&0.2617&0.1471&0.1575&0.155&0.0851&0.0832&0.0791&0.0679&0.0661&0.0625\\
			\hline
			FM&0.2756&0.2836&0.2817&0.1690&0.1691&0.1655&0.0973&0.0997&0.0921&0.0770&0.0780&0.0728\\
			\hline
			TrustSVD&0.2913&0.2880&0.2915&0.1754&0.1723&0.1738&0.1372&0.1367&0.1427&0.1062&0.1047&0.1085\\
			\hline
			ContextMF&0.2985&0.3011&0.3043&0.1788&0.1808&0.1818&0.1217&0.1201&0.1265&0.0963&0.0943&0.0961\\
			\hline
			PinSage&0.2952&0.2958&0.3065&0.1758&0.1779&0.1868&0.1209&0.1227&0.1142&0.0952&0.0978&0.0991\\
			\hline
			SocialGCN&\textbf{0.3283}&\textbf{0.3360}&\textbf{0.3364}&\textbf{0.1978}&\textbf{0.2020}&\textbf{0.2023}&\textbf{0.1575}&\textbf{0.1621}&\textbf{0.1594}&\textbf{0.1210}&\textbf{0.1231}&\textbf{0.1234}\\
			\hline
		\end{tabular}
	\vspace{-0.5cm}
	\end{small}
\end{table*}

\begin{table*}
	\begin{small}
		\centering
		\caption{HR@N and NDCG@N comparisons for different top-N values.}\label{tab:hr_ndcg_topk}
		\begin{tabular}{|c|c|c|c|c|c|c|c|c|c|c|c|c|}
			\hline
			\multirow{3}*{Models} &\multicolumn{6}{|c|}{\textit{Yelp}}&\multicolumn{6}{|c|}{\textit{Flickr}} \\
			\cline{2-13}
			&\multicolumn{3}{|c|}{HR}&\multicolumn{3}{|c|}{NDCG}&\multicolumn{3}{|c|}{HR}&\multicolumn{3}{|c|}{NDCG}\\
			\cline{2-13}
			&N=5&N=10&N=15&N=5&N=10&N=15&N=5&N=10&N=15&N=5&N=10&N=15\\
			\hline
			\hline
			BPR&0.1713&0.2632&0.3289&0.1243&0.1575&0.1773&0.0657&0.0851&0.1041&0.0607&0.0679&0.0737\\
			\hline
			FM&0.1832&0.2836&0.3485&0.1343&0.1691&0.1898&0.0705&0.0997&0.1191&0.0633&0.0780&0.0817\\
			\hline
			TrustSVD&0.1906&0.2915&0.3693&0.1385&0.1754&0.1983&0.1072&0.1427&0.1741&0.0970&0.1085&0.1200\\
			\hline
			ContextMF&0.2045&0.3043&0.3832&0.1484&0.1818&0.2081&0.0928&0.1265&0.1637&0.0823&0.0963&0.1091\\
			\hline
			PinSage&0.2099&0.3065&0.3873&0.1536&0.1868&0.2130&0.0925&0.1227&0.1489&0.0842&0.0991&0.1036\\
			\hline
			SocialGCN&\textbf{0.2162}&\textbf{0.3364}&\textbf{0.4041}&\textbf{0.1601}&\textbf{0.2023}&\textbf{0.2226}&\textbf{0.1210}&\textbf{0.1621}&\textbf{0.1961}&\textbf{0.1085}&\textbf{0.1234}&\textbf{0.1341}\\
			\hline
		\end{tabular}
	\vspace{-0.5cm}
	\end{small}
\end{table*}

\textbf{Datasets.} 
\textit{Yelp} is an online location-based social network. Users make friends with others and express their experience through the form of reviews and ratings. As each user give ratings in the range $[0,5]$, similar to many works,  we transform the ratings that are larger than 3 as the liked items by this user. As the rich reviews are associated with users and items, we use the
popular gensim tool\footnote{https://radimrehurek.com/gensim/} to learn the embedding of each word with Word2vec model~\cite{mikolov2013distributed}. Then, we get the feature vector of each user~(item) by averaging all the learned word vectors of the user(item).

\textit{Flickr} is a who-trust-whom online image based social sharing platform. Users follow other users and share their preferences to images to their social followers. Users express their preferences through the upvote behavior. For research purpose, we crawl a large dataset from this platform. Given each image, we have a ground truth classification of this image on the dataset. We send images to a VGG16 convolutional neural network and treat the 4096 dimensional representation in the last connected layer in VGG16 as the feature representation of the image~\cite{simonyan2014very}. For each user, her feature representation is the average of the image feature representations she liked in the training data.

In the data preprocessing step, for both datasets, we filtered out users that have less than 2 rating records and 2 social links. And removed the items which have been rated less than 2 times. We randomly select 10\% of the data for the test. In the remaining 90\% data, to tune the parameters, we select 10\% from the training data as the validation set. The detailed statistics of the data after preprocessing is shown in Table~\ref{tab:stat}.

\textbf{Baselines and Evaluation Metrics.}
We compare SocialGCN with various state-of-the-art baselines. The details of these baselines are listed as follows:

\begin{itemize}
	\vspace{-0.1cm}
	\item \textbf{BPR} It is a competing latent factor model for implicit feedback based recommendation. It designed a ranking based function that assumes users prefer items they like compared to unobserved ones.\cite{UAI2009bpr}.
	\vspace{-0.1cm}
	\item \textbf{FM}  This model is a unified latent factor based model that leverages the user and item attributes. In practice, we use the user and item features as introduced above\cite{ICDM2010factorization}.
	\vspace{-0.1cm}
	\item \textbf{TrustSVD}  This model incorporates the trust influence from social neighbors on top of SVD++. It shows state-of-the-art performance on social recommendation results\cite{guo2015trustsvd}.
	\vspace{-0.1cm}
	\item \textbf{ContextMF} This method combines social context and social network under a collective matrix factorization framework with carefully designed regularization terms\cite{TKDE2014scalable}. We use the user and item features as the context information.
	\vspace{-0.1cm}
	\item \textbf{PinSage} It is a state-of-the-art model for designing efficient convolutional operations for web-scale recommendations. As the original PinSage focuses on generating high-quality embeddings of items, we generalize this model by constructing a user-item bipartite for recommendation~\cite{KDD2018graph}.
\end{itemize}

\begin{table*}
	\centering
	\caption{HR@10 and NDCG@10 of our simplified models on \textit{Yelp} and \textit{Flickr}. $K$=1 denotes we do not consider information diffusion, \textbf{X}=\textbf{Y}=0 denotes the user and item feature vector is not available, and \textbf{P}=0 denotes we do not add the free base user latent vector.}\label{tab:submodels_y}
	\begin{tabular}{|c|c|c|c|c|c|c|c|c|}
		\hline
		\multirow{2}*{Simplified models} &\multicolumn{4}{|c|}{\textit{Yelp}}&\multicolumn{4}{|c|}{\textit{Flickr}}\\
		\cline{2-9}
		&HR&Improve.&NDCG&Improve.&HR&Improve.&NDCG&Improve.\\
		\hline
		SocialGCN&\textbf{0.3364}&-&\textbf{0.2023}&-&\textbf{0.1621}&-&\textbf{0.1231}&-\\
		\hline
		SocialGCN($K$=1)&0.3280&-2.50\%&0.1984&-1.93\%&0.1573&-2.96\%&0.1216&-1.22\%\\
		\hline
		SocialGCN(\textbf{X}=\textbf{Y}=0,$K$=2)&0.3218&-4.34\%&0.1915&-5.34\%&0.1586&-2.16\%&0.1217&-1.14\%\\
		\hline
		SocialGCN(\textbf{X}=\textbf{Y}=0,$K$=1)&0.3181&-5.44\%&0.1869&-7.61\%&0.1407&-13.20\%&0.1075&-12.67\%\\
		\hline
		SocialGCN(\textbf{P}=0)&0.2381&-29.22\%&0.1496&-16.05\%&0.1043&-35.68\%&0.0835&-32.17\%\\
		\hline
	\end{tabular}
\vspace{-0.5cm}
\end{table*}

As we focus on recommending top-N items for each user, we use two widely adopted ranking based metrics: Hit Ratio~(HR) and Normalized Discounted Cumulative Gain(NDCG)~\cite{SIGIR2018attentive}. Specifically, HR measures the number of items that the user likes in the test data that has been successfully predicted in the top-N ranking list. And NDCG considers the hit positions of the items and gives a higher score if the hit items in the top positions.
For both metrics, the larger the values, the better the performance.
Since there are too many unrated items, in order to reduce the computational cost, for each user, we randomly sample 1000 unrated items at each time and combine them with the positive items the user likes in the ranking process. We repeat this procedure 10 times and report the average ranking results.


\textbf{Parameter Setting.} For all the models that are based on the latent factor models, we initialize the latent vectors with small random values. In the model learning process, we use Adam as the optimizing method for all models that relied on the gradient descent based methods with a learning rate of 0.001. And the batch size is set as 512. In our proposed SocialGCN model, we set the regularization parameter as $\lambda$=0.0001. For the aggregation function in the convolutional operation, we have tried the max pooling and average pooling. We find the average pooling usually shows better performance. Hence, we set the average pooling as the aggregation function. Similar to many GCN models~\cite{KDD2018graph,ICLR2017semi}, we set the depth parameter $K$=2. We use $ReLU$ to implement the non-linear transformation function $\sigma$ in ~(Eq.\eqref{eq:item_embed}) and ~(Eq.\eqref{eq:base_user}). There are several other parameters in the baselines, we tune all these parameters to ensure the best performance of the baselines for fair comparison. Please note that as generating user and item features are not the focus of our paper, we use the feature construction techniques as mentioned above.

\vspace{-0.4cm}
\subsection{Overall Comparison}
\vspace{-0.1cm}
In this section, we compare the overall performance of all models on two datasets. Specifically, Table \ref{tab:hr_ndcg_d} shows the HR@10 and NDCG@10 results for both datasets with varying latent dimension size $D$. As can be seen from this table, on both datasets, our model consistently outperforms all the other models with different values of $D$ for the two ranking metrics. E.g, when $D$=64, the improvement of HR(NDCG) over the best baselines is 11.70\%(13.73\%) on \textit{Flickr}. Among all the baselines, BPR only considered the user-item rating information for recommendation FM and TrustSVD improve over BPR by leveraging the node features and social network information. PinSage takes the same kind of input as FM and shows better performance than FM, showing the effectiveness of GCN. When comparing the results of the two datasets, we observe that leveraging social network contributes more on \textit{Flickr} compared to \textit{Yelp}. We guess a possible reason is that, as shown in Table\ref{tab:stat}, \textit{Flickr} dataset is much sparser than \textit{Yelp}. Therefore, the social network could alleviate the data sparsity issue in \textit{Flickr} to some extent. Last but not least,  we find the performance does not increase as the latent dimension size $D$ increases from 16 to 64. 
In the following experiment, we set the proper $D$ for each model with the best performance in order to ensure fairness.

Table \ref{tab:hr_ndcg_topk} shows the HR@N and NDCG@N on both datasets with varying top-N recommendation size N. From the results, we also find similar observations as Table \ref{tab:hr_ndcg_d}, with our proposed model SocialGCN always shows the best performance. Based on the overall experiment results, we could empirically conclude that our proposed SocialGCN model outperforms all the baselines under different ranking metrics and different parameters.

\vspace{-0.2cm}
\subsection{Detailed Model Analysis}
\vspace{-0.1cm}
In this subsection, we would like to give a detailed analysis of our proposed model and show the effectiveness of each part in SocialGCN. As shown in the model part, there are three characteristics in the modeling process: the social diffusion with depth $K$~(Eq.\eqref{eq:pro_gcn}), the embeddings  $\mathbf{h}^0_a$ of each user $a$ that incorporates the free embedding $\mathbf{x}_a$  and feature vector $\mathbf{p}_a$~(Eq.\eqref{eq:base_user}). When $K$ equals 1, our model degenerates to a social recommendation model that only considers the neighborhood information without the social diffusion process. Therefore, we would like to show the effectiveness of the social diffusion compared to $K$=1, the effectiveness of the free embedding compared to $\mathbf{p}_a$=0, and the effectiveness of the user features compared to $\mathbf{x}_a$=0.

In Table~\ref{tab:submodels_y}, we have
listed the simplified variants of our proposed SocialGCN model. The Improve. represents the comparison between the performance of current model with SocialGCN. As can be seen from this table, as we do not consider the diffusion process~(\textit{K}=1), the recommendation performance drops. This situation becomes more severe when the user and item attributes are not available~(i.e., \textbf{X}=\textbf{Y}=0). E.g., the HR@10 drops 11.29\% on the \textit{Flickr}. We also notice that it is very important to add the free base latent vector of users and items in the modeling process, as the social network and the feature could not well model the complete latent factors of users and items.  Therefore, all the proposed parts are important in SocialGCN for recommendation performance.

\vspace{-0.4cm}
\section{Conclusions}
\vspace{-0.2cm}
In this paper, we proposed a SocialGCN model for social recommendation problem. 
Our model combines the strengths of GCNs for modeling the diffusion process in social networks and the classical latent factor based models for capturing user-item preferences.  Specifically, the user embeddings are built in a layer-wise diffusion manner, with the initial user embedding as a function of the current user's features and a free base user latent vector that is not contained in the user feature vector.  Similarly, each item's latent vector is also a combination of the item's free latent vector, as well as its feature representation.
We showed that the proposed SocialGCN model is flexible when the user and item attributes are not available. The experimental results clearly showed the flexibility and effectiveness of our proposed models. E.g., SocialGCN improves 13.73\% over the best baseline of NDCG on \textit{Flickr}. In the future, we would like to explore GCNs for more social recommendation applications, such as social influence modeling, temporal social recommendation, and so on.

\begin{small}
\bibliographystyle{aaai}
\bibliography{aaai2019}
\end{small}

\end{document}